\documentclass[aps,prl,preprintnumbers,english,footinbib,floatfix,reprint,twocolumn,showpacs,amsmath,amssymb,superscriptaddress,longbibliography,10pt,tightenlines]{revtex4-2}

\usepackage{graphicx, epsf,xcolor}% Include figure files
\newcommand{\br}{{\bf r}}

\newcommand{\ff}{{\mathbf{f}}}

\newcommand{\bff}{{\bf f}}

\newcommand{\expv}[1]{\big\langle #1 \big\rangle}
\newcommand{\Expv}[1]{\Big\langle #1 \Big\rangle}

\usepackage{subcaption}
\usepackage{ragged2e}
\DeclareCaptionJustification{justified}{\justifying}
\usepackage{physics}
\usepackage{hyperref}
\hypersetup{
    colorlinks=true,
    linkcolor=blue,
    filecolor=magenta,      
    urlcolor=cyan,
    }

\begin{document}

\title{Finite temperature correlation functions of the sine--Gordon model}
\author{M. T\'oth}
\affiliation{Department of Theoretical Physics, Institute of
Physics, Budapest University of Technology and
Economics, M{\H u}egyetem rkp. 3., H-1111 Budapest,
Hungary}
\affiliation{BME-MTA Statistical Field Theory `Lend\"ulet’ Research
Group, Budapest University of Technology and
Economics, M{\H u}egyetem rkp. 3., H-1111 Budapest,
Hungary}
\author{J. H. Pixley}
\affiliation{Department of Physics and Astronomy, Center for Materials Theory, Rutgers University, Piscataway, New Jersey 08854, USA}
\affiliation{Center for Computational Quantum Physics, Flatiron Institute, 162 5th Avenue, New York, NY 10010}
\author{G. Tak\'acs}
\affiliation{Department of Theoretical Physics, Institute of
Physics, Budapest University of Technology and
Economics, M{\H u}egyetem rkp. 3., H-1111 Budapest,
Hungary}
\affiliation{BME-MTA Statistical Field Theory `Lend\"ulet’ Research
Group, Budapest University of Technology and
Economics, M{\H u}egyetem rkp. 3., H-1111 Budapest,
Hungary}
\author{M. Kormos}
\affiliation{Department of Theoretical Physics, Institute of
Physics, Budapest University of Technology and
Economics, M{\H u}egyetem rkp. 3., H-1111 Budapest,
Hungary}
\affiliation{BME-MTA Statistical Field Theory `Lend\"ulet’ Research
Group, Budapest University of Technology and
Economics, M{\H u}egyetem rkp. 3., H-1111 Budapest,
Hungary}
\affiliation{HUN-REN-BME-BCE Quantum Technology Research Group,
Budapest University of Technology and Economics, M{\H u}egyetem rkp. 3., H-1111 Budapest, Hungary}

\date{10th April 2026}

\begin{abstract}
The sine--Gordon model serves as a foundational $1+1$-dimensional quantum field theory with numerous applications in condensed matter physics. Despite its integrability, characterizing its finite-temperature behavior remains a significant theoretical challenge. Here we use the previously developed Method of Random Surfaces (MRS) to evaluate two-point and higher-order correlation functions. We cross-check these results with known analytical limits, demonstrating that the MRS provides reliable, non-perturbative data in intermediate regimes where traditional form-factor expansions and semiclassical methods are inapplicable. Furthermore, we derive an exact result for arbitrary $N$-point functions satisfying an appropriate selection rule, providing a direct computational method for complex multi-point observables at finite temperature. We also characterize the non-Gaussianity of correlations and demonstrate that the results align with intuitive theoretical expectations. \end{abstract}

\maketitle

\paragraph{Introduction.---} The sine--Gordon (sG) model is a paradigmatic $1+1$-dimensional quantum field theory that can be used to describe a wide range of condensed matter systems \cite{tsvelik_2003,Giamarchi:743140}. Its applications as a low-energy effective theory range from  trapped ultra-cold atoms \cite{2007PhRvB..75q4511G,2010PhRvL.105s0403C,2010Natur.466..597H,2017Natur.545..323S,2024PhRvB.109c5118B} through quasi-1D antiferromagnets, carbon nanotubes and organic conductors \cite{Controzzi2001,2005ffsc.book..684E}
to quantum circuits \cite{2021NuPhB.96815445R} and coupled spin chains \cite{Wybo2022,PRXQuantum.4.030308}. It is a prime example of a quantum field theory with a strong-weak coupling duality \cite{1975PhRvD..11.2088C,1975PhRvD..11.3026M} and non-trivial non-perturbative dynamics. The sG model is integrable, which provides access to exact, nonperturbative results including its exact $S$-matrix \cite{1977CMaPh..55..183Z,ZAMOLODCHIKOV1979253}, 
its thermodynamics via the nonlinear integral equation 
\cite{1991JPhA...24.3111K,1995NuPhB.438..413D,Hegedus2025,Hegedus2026} and the thermodynamic Bethe ansatz \cite{koch2023exact,2024ScPP...16..145N}, as well as its form factors \cite{1992ASMP...14....1S} and exact zero-temperature expectation values of local operators \cite{Lukyanov1996}.

Despite the integrability of the model, describing its behavior at finite temperature remains a formidable challenge. 
Traditional analytical and numerical approaches suitable for zero temperature, such as form-factor expansions \cite{2005ffsc.book..684E,2014JHEP...03..026B,2009JSMTE..09..018E,2010JSMTE..11..012P} or truncated Hamiltonian approaches \cite{Yurov:1989yu,1998PhLB..430..264F,2018PhRvL.121k0402K,2018RPPh...81d6002J,2022CoPhC.27708376H,2020arXiv200513544A}, face serious difficulties at finite temperature. Approximate semiclassical methods \cite{2005PhRvL..95r7201D,2017PhRvL.119j0603M,PhysRevB.106.205151} are restricted to
very low temperatures.
%often restricted to specific limiting regimes. 
Ballistic fluctuation theory \cite{Myers2020,Doyon2019b,DelVecchio2023} based on generalized hydrodynamics cannot access the correlation functions of the vertex operators studied in this work.

Consequently, there is currently no general approach for obtaining the theory’s correlation functions at finite temperature. In a recent work \cite{Toth2025}, we demonstrated that the method of random surfaces (MRS) \cite{2008PhRvA..77f3606I,2008NatPh...4..489H,2010PhRvA..82c2118O,2013PhRvB..87a4305O} offers a powerful alternative, accurately computing the free energy density and expectation values of exponential operators. In this Letter, we extend the MRS framework to calculate the two-point and higher-order correlation functions of the sine--Gordon model at finite temperature. Unlike existing methods, the MRS provides the first reliable results for correlation functions at intermediate values where neither classical nor form-factor expansions are fully applicable.  We demonstrate that the MRS can be used to derive exact results for arbitrary multipoint functions of vertex operators. By benchmarking our numerical results in various limiting cases, we show that the correlation length converges to the expected conformal result in the high-temperature limit and to the mass of the first breather in the low-temperature limit. Furthermore, we investigate the impact of finite-size effects and numerical artifacts, such as Fourier mode truncation, to establish the reliability of the approach. Our results suggest that the MRS is a robust tool for exploring the finite-temperature dynamics of $(1+1)D$ field theories.

\begin{figure}[t!]
    \centering
    \includegraphics[width=0.6\columnwidth]{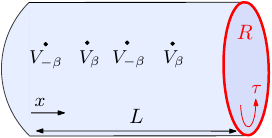}
    \includegraphics[width=0.75\columnwidth]{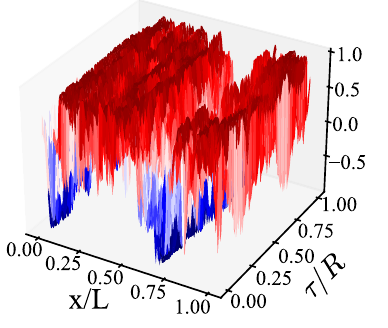}
    \caption{\emph{Top:} The system we consider lives on a cylinder of length $L$ (coordinate $x$) and circumference $R$ (coordinate $\tau)$. We depict  4 vertex operators  $V_{\pm \beta}$ inserted for an equal-time correlation function. \emph{Bottom:} Real part of $h(\{t_\ff\}, \br)$ in Eq.\ \eqref{eq:hequation} for a set of random $t_\ff$ values, representing a typical `random surface' we are integrating over.}
    \label{fig:randomsurface}
\end{figure}

%\par\textit{Finite temperature correlation functions} -- 
\paragraph{Finite-temperature correlation functions.---} 
The sine--Gordon model is a $1+1$ dimensional field theory governed by the Euclidean action
\begin{equation}
    S=\int_{-\infty}^\infty dx\int_0^R d\tau\frac{1}{2}
    \left[(\partial_{\tau} \phi)^2 +  (\partial_{x} \phi)^2\right] - \lambda_0 \cos(\beta\phi)
\end{equation}
at finite temperature $T=1/R$ 
%($R$ denotes the ``radius'' of the cylinder),
thus the imaginary time $\tau$ satisfies $0\leq\tau\leq R$, and $\phi(x,\tau)$ is a real scalar field living on a cylinder of circumference $R$, see Fig.~\ref{fig:randomsurface}. We calculate the higher-order correlation functions of vertex operators
\begin{equation}
    \hat V_\alpha(\br) = a^{-\frac{\alpha^2}{4\pi}} e^{i\alpha\hat\phi(\br)},
    \label{eq:vertexoperator}
\end{equation}
where $\br=(x,\tau)$ and $a$ is a UV regulator length.
%introduced in the Green's function. 
We first focus on the two-point function of such operators, which we can get directly from the partition function using functional derivative methods (See Supplemental Material \cite{SM} for details.). The partition function can be written as a product $Z=Z_0Z_I$ of the free bosonic partition function $Z_0=\int D[\phi]\, e^{-S_0}$ and an interacting part
\begin{align}
    Z_I = \displaystyle\frac{\int D[\phi]\, e^{-S_0-S_I}}{\int D[\phi]\, e^{-S_0}} 
    \equiv \langle e^{-S_I} \rangle_0 \,,
    \label{eqn:avg0}
\end{align}
where $S_0$ is the free ($\lambda=0$) and $S_I=-\lambda \int dxd\tau \cos(\beta\phi)$ is the interacting part of the sine--Gordon action. For the functional derivative approach, we promote the coupling strength to be space-time dependent:
%rewrite the interaction in the action as
\begin{equation}
    S_I = \lambda\left(\frac{\sigma_1(\br)}{2}\hat V_\beta(\br_1) + \frac{\sigma_2(\br)}{2}\hat V_{-\beta}(\br_2)\right)\,,
\end{equation}
where $\lambda=\lambda_0 a^{\beta^2/4\pi}$. This allows us to calculate the lowest non-trivial (neutral) correlator as
\begin{equation}
    \expv{ \hat V_{\beta}(\br_1) \hat V_{-\beta}(\br_2)} = 4\left.\frac{\delta^2 Z_I}{\delta \sigma_1(\br_1) \delta \sigma_2(\br_2)}\right|_{\sigma_1(\br) = \sigma_2(\br) = 1}.
    \label{eqn:funder}
\end{equation}
Generalizing the derivation of Ref.\ \cite{Toth2025} to this case, the partition function is
\begin{equation}
    Z_I = \int \mathcal{D}[t_\ff] I_0\left(\lambda R^{-\Delta}\sqrt{g_1(\{t_\ff\}) g_2(\{-t_\ff\})}\right)\,,
\label{eq:partfuncdiffg}
\end{equation}
where $I_0(x)$ is the modified Bessel function of the first kind, $\Delta=\beta^2/(4\pi)$, and $g_j(\{t_\ff\}) = \int d\br \sigma_j(\br) h(\{t_\ff\},\br)$ with
\begin{equation}
    h(\{t_\ff\},\br) = C
%    \frac{\lambda_j(\br)}{\lambda}
    \exp\left\{ \sum_\ff^{m_{\mathrm{max}}}  it_\ff\sqrt{\Delta G_\ff}\,\psi_\ff(\br) 
    \right\},
\label{eq:hequation}
\end{equation}
where $C$ is a constant prefactor and $G_\ff$ and $\psi_\ff$ appear in the projector decomposition of the non-interacting Green's function:
$G(\br_1,\br_2) = \sum_\ff G_\ff \psi_\ff(\br_1)\psi_{\ff}(\br_2)$,
which is achieved by a Fourier series expansion in $x$ and $\tau$.

We used the shorthand notation $\int \mathcal{D}[t_\ff] = \prod_\ff \int_{-\infty}^\infty \frac{dt_\ff}{\sqrt{2\pi}} e^{-t_\ff^2/2}$ for the Gaussian integrals introduced for every Fourier mode. In practice, we evaluate this high dimensional integral using a Monte Carlo approach: we draw a set of random numbers $\{t_\ff\}$ from the Gaussian distribution with zero mean and unit variance, which makes $h(\{t_\ff\},\br)$ a random function of $\br=(x,\tau)$, and evaluate the 2D integrals over the cylinder for $g_j(\{t_\ff\})$ (for a random surface such as Fig.\ \ref{fig:randomsurface}), and average $I_0(\dots)$ over different sets of $\{t_\ff\}$.
From Eqs. \eqref{eqn:funder} and \eqref{eq:partfuncdiffg} we obtain 
%which yields
\begin{widetext}
\begin{equation}
\label{eq:twopoint}
    \expv{ \hat V_{\beta}(\br_1) \hat V_{-\beta}(\br_2)} = \frac{R^{-2\Delta}}{Z_I(\lambda)}\int \mathcal{D}[t_\ff]  I_0\left(\lambda R^{-\Delta} \sqrt{g(\{t_\ff\}) g(\{-t_\ff\})}\right) h(\{t_\ff\}, \br_1) h(\{-t_\ff\}, \br_2)\, .
\end{equation}
\end{widetext}

\begin{figure*}[t!]
    \centering
    \includegraphics[width=\textwidth]{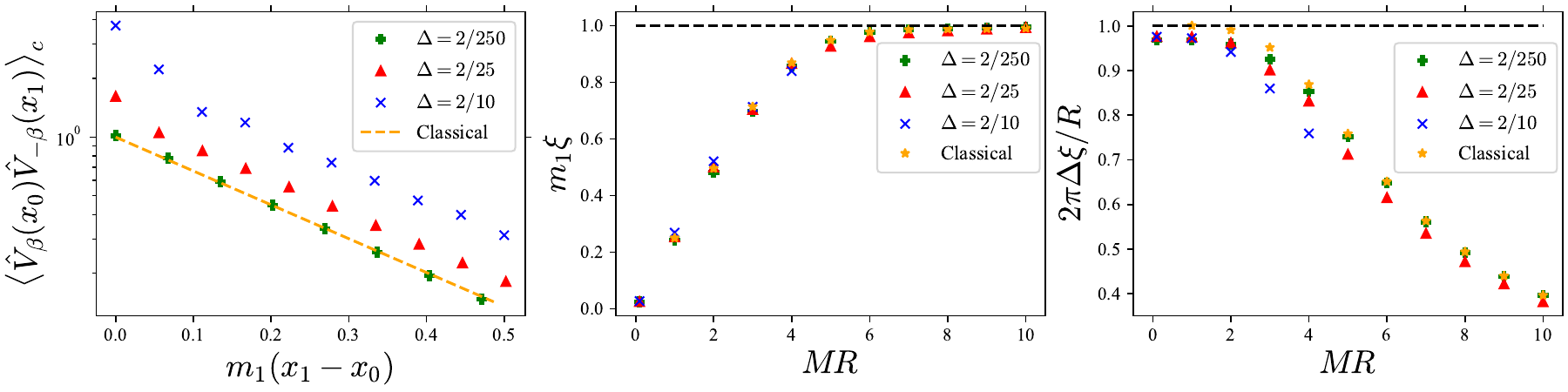}
    \caption{\emph{Left:} Equal-time connected correlation functions of vertex operators for three different couplings $\Delta=2/250$, $\Delta=2/25$, and $\Delta=2/10$ at inverse temperature $MR=1$. The dashed orange line is the numerical result for the classical sine-Gordon model \cite{koch2023exact}. 
    %The correlation lengths are similar for all couplings but 
    The smallest coupling is in the semiclassical regime, but for larger couplings the correlator deviates from the classical result. The results were obtained at a fixed maximal mode number $m_{max}=60$ using $\sim 1.5\cdot10^6$ Monte Carlo samples; the statistical errors are smaller than the symbol size. \emph{Middle:} Correlation length against the dimensionless inverse temperature in units of the inverse mass gap $m_1^{-1}$. For low temperature, the correlation lengths approach the inverse mass gap, and show very weak dependence on the coupling strength.
    %at higher coupling strengths, the low-temperature evaluations were too noisy to yield useful results. 
    \emph{Right:} Correlation length vs.\ the dimensionless inverse temperature rescaled with the expected infinite temperature value $R/(2\pi\Delta)=2R/\beta^2$.
    }
    \label{fig:corrfunc}
\end{figure*}

Note that the parameter $\lambda$ is related to the soliton mass via the relation 
%
%\begin{equation}
$\lambda = \kappa(\Delta) M^{2-\Delta}$,
%\end{equation}
%
where $\kappa(\Delta)$ is exactly known \cite{1995IJMPA..10.1125Z}. This in turn sets the dimensionless temperature $MR$.

The MRS calculation introduces a few numerical artifacts. As we need a discrete Fourier expansion, we restricted the integration along the $x$ direction to the interval $[-L/2,L/2]$ and made the Green's function periodic on $[-L,L]$, so we can integrate over a finite cylinder while using the Green's function of the infinite one. However, this restriction may introduce potential finite-size effects. 
In addition, the number of Fourier modes is truncated to a finite value $m_\text{max}$ which is equivalent of an ultraviolet cut-off~\cite{Toth2025}. Finally, the integral of \eqref{eq:hequation} is calculated numerically by discretizing the $\psi_\ff$ functions on a grid.

\par The numerical results for the equal-time connected correlation function are shown in Fig.\ \ref{fig:corrfunc} on logarithmic scale for various coupling strengths $\Delta=\beta^2/(4\pi)$. The lowest coupling strength $\Delta=2/250$ apparently lies in the semiclassical regime as the correlator for this coupling coincides with the classical result \cite{koch2023exact}. However, for larger couplings, the correlation function deviates from the classical result and depends on the coupling strength, which is a genuine quantum effect. 
%While the correlation length characteristic of the asymptotic exponential decay is similar for all couplings, the correlation functions themselves vary with $\Delta$. In particular, }.
As in our previous results \cite{Toth2025}, the method performs best in the high-temperature regime $MR\lesssim5$. 
%The partition function and the expectation values $Z_I\cdot\langle.\rangle$ grow with the dimensionless inverse temperature through the defining Bessel function, which introduces larger numerical errors.
\par Although we cannot compare the correlation function to any pre-existing result, we can extract the correlation length characterizing the asymptotic exponential decay and benchmark it using certain limiting cases. In the low-temperature limit, it is expected to approach the inverse mass gap $m_1$ given by the mass of the lightest breather,
%\cite{2020JHEP...07..224K} 
while in the high-temperature limit, we should recover the conformal result,  $2R/\beta^2$ \cite{tsvelik_2003}. These two limits are illustrated in the middle and right panels of Fig.\ \ref{fig:corrfunc}.
%\mk{These plots also show genuine quantum results for couplings corresponding to $\Delta=2/10$ and $\Delta=2/25$.
%with results obtained in the quantum regime as indicated 
%For these couplings, the vertex-operator expectation values
%When cross-checking the numerical data for these expectation values against those obtained for the 
%were found to differ significantly from those calculated in the classical sine–Gordon model \cite{Toth2025}.
%we observed noticeable discrepancies which decrease and eventually vanish as $\Delta$ is reduced. 
Interestingly, the correlation length is found to depend rather weakly on the coupling strength.
%shows good agreement between the MRS results in this 
%regime and the 
%numerically close to the corresponding classical predictions.} 

\begin{figure*}
    \centering
    \includegraphics[width=\textwidth]{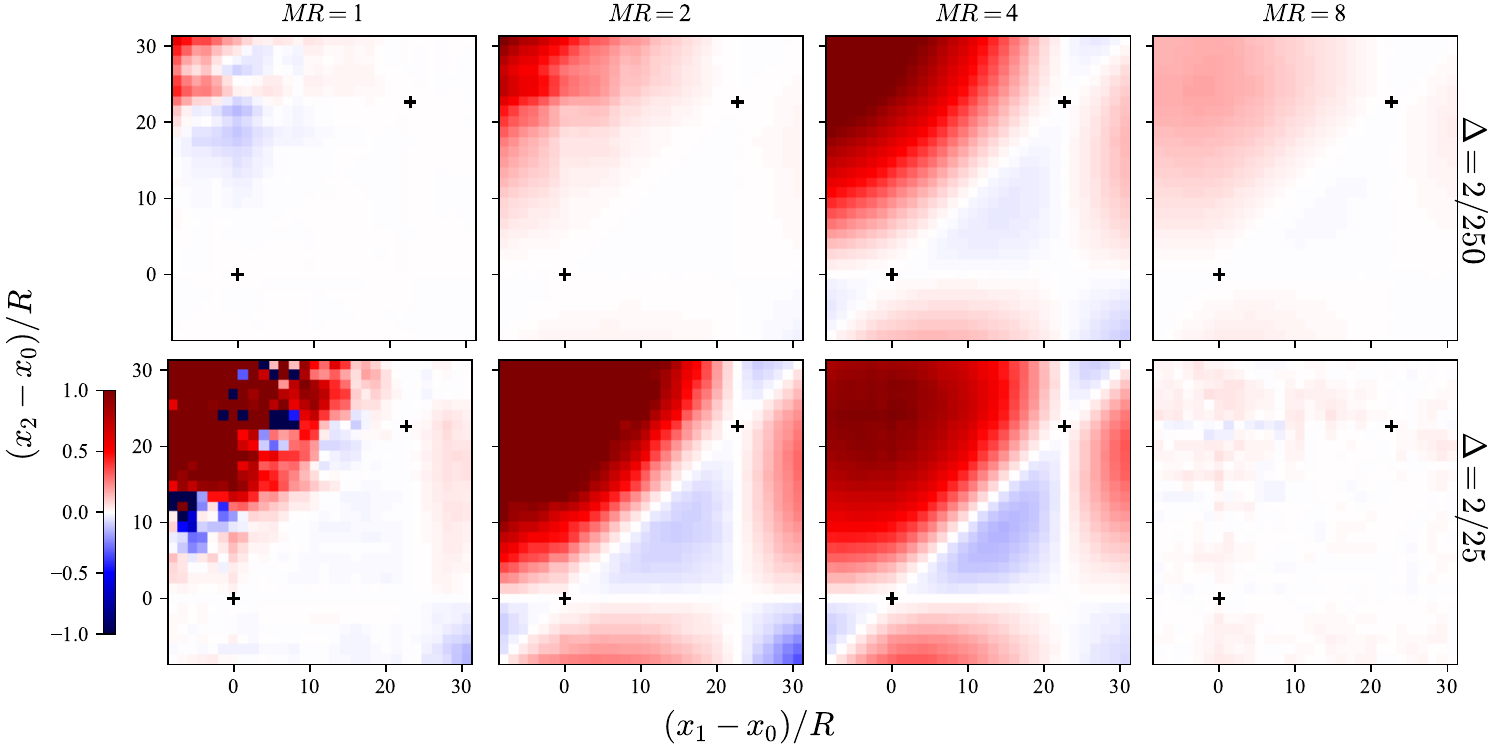}
    \caption{Four-point function $K_4((\{x_i\})$ defined in Eq.\ \eqref{eq:fourpointquantity} for weak (top) and intermediate (bottom) coupling and different inverse temperature values. To visualize the data, we fixed the location of 2 operators, $x_0$ and $x_3$ at $0.39L$ and $0.61L$ respectively, and swept the available interval with the other two (the fixed positions are represented with black crosses). Data in each row is normalized by the highest value in the third column. We observe the strongest correlations in the intermediate temperature regime.}
    \label{fig:fourpoint}
\end{figure*}

\paragraph{Multipoint correlation functions.---} Higher order correlation functions contain essential information about the physics of the sine--Gordon model. Therefore, it would be beneficial to be able to calculate these quantities; however, the available methods are even more restricted than for two-point functions, and there is essentially no efficient and accurate method to obtain them. Fortunately, the MRS helps us overcome these longstanding limitations. We aim to calculate the higher-order correlation functions of vertex operators \eqref{eq:vertexoperator}
\begin{equation}
   C_{\{\alpha_i\}}(\br_1, ...,\br_N) = \expv{ \prod_i^N \hat V_{\alpha_i}(\br_i)}\,.
\end{equation}
%with the exact form of the vertex operators defined in \eqref{eq:vertexoperator}. 
In the case when the different $\alpha$ values are equal to $\pm \beta$, 
%the representation of the correlation function can be built 
the result can be derived 
using functional derivative methods, similarly to the two-point function. However, we can obtain an exact result for arbitrary multipoint functions of such a form, even with $\alpha$ values that are different from $\pm\beta$, provided they satisfy the selection rule
\begin{equation}
    \sum_i \alpha_i = n\beta,\quad n \in \mathbb{Z}\,.
\end{equation}
Using the approach of Ref.\ \cite{Toth2025} yields the result
\begin{widetext}
\begin{equation}
    C_{\{\alpha_i\}}(\br_1, ...,\br_N) = \frac{R^{-\Delta_\Sigma}}{Z_I(\lambda)} \int \mathcal{D}[t_\ff]\, 
    \left(\frac{g(\{-t_\ff\})}{g(\{t_\ff\})}\right)^{\frac s{2}} \prod_i
    h_{\alpha_i}(\{t_\ff\}, \br_i)\,
    I_{\left|s\right|}\left( \lambda R^{-\Delta} \sqrt{g(\{t_\ff\})g(\{-t_\ff\})}\right),
\end{equation}
\end{widetext}
with 
\begin{equation}
h_{\alpha_i}(\{t_\ff\}, \br_i) = C^{s^2_i} \exp(\sum_\ff\left[ i t_\ff s_i\sqrt{\Delta G_\ff} \psi_\ff(\br_i) \right]),
\end{equation}
$\Delta_\Sigma = \sum_i\frac{\alpha_i^2}{4\pi}$, $s_i = \alpha_i/\beta$ and $s = \sum_i s_i$ {(see End Matter for details)}. In the special case $\{\alpha_i\} = \{\beta, -\beta\}$ we recover the result \eqref{eq:twopoint} for the two-point function. 
\par
\begin{figure}[t!]
    \centering
    \includegraphics[width=\columnwidth]{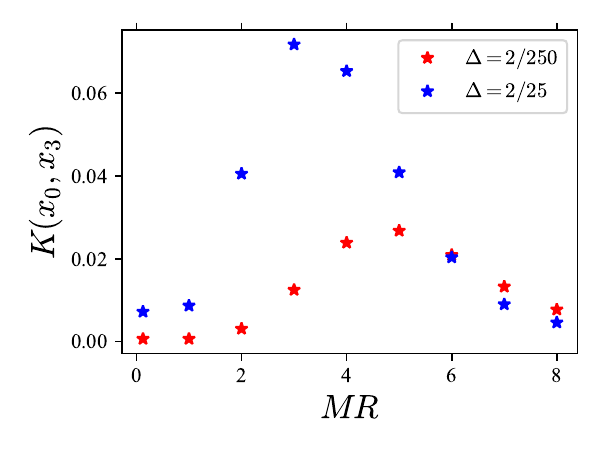}
    \caption{The kurtosis-like quantity $K(x_0,x_3)$ defined in Eq.\ \eqref{eq:kurtosis} as a function of the inverse temperature for two different couplings. The two fixed spatial coordinates are  $x_0=0.39L$ and $x_3=0.61L$, the other two are integrated over the interval $[0.3L, 0.7L]$. For both couplings, the measure decreases in both the low- and high-temperature regimes, indicating almost Gaussian correlations, while in the intermediate regime, we observe strong non-Gaussianity, with a strength increasing with the coupling. The errors are smaller than the symbol size.}
    \label{fig:kurtosis}
\end{figure}

 To represent higher order correlation functions, we calculated the equal-time 4-point function $C_{\{\beta, \beta, -\beta,-\beta\}}(x_0, x_1, x_2, x_3)$ at $\mathbf{r_j}=(x_j,0)$ (denoted $C_4(\{x_i\})$ in the following). For a free (Gaussian) scalar field for which the only non-vanishing connected correlation function is the two-point function, evaluating the 4-point function using Wick's theorem results in the product form
\begin{equation}
    C^\text{Gauss}_4(\{x_i\}) = \frac{C_{+-}(0,2)C_{+-}(0,3)C_{+-}(1,2)C_{+-}(1,3)}{C_{+-}(0,1)C_{+-}(2,3)}
%  C^\text{Gauss}_4(\{x_i\}) = \frac{C_{+-}(0,2)C_{+-}(\mathbf{0},\mathbf{3})C_{+-}(\mathbf{1},\mathbf{2})C_{+-}(\mathbf{1},\mathbf{3})}{C_{+-}(\mathbf{0},\mathbf{1})C_{+-}(\mathbf{2},\mathbf{3})}
    \label{eq:correct_disconnected}
\end{equation}
with $C_{+-}(i,j) = \expv{\hat V_{\beta}(x_i)\hat V_{-\beta}(x_j)}$ {(c.f. End Matter)}.
Therefore, the effect of interactions can be characterized by the connected part of the 4-point function, which can be quantified by the quantity 
\begin{equation}
    K_4((\{x_i\})=\frac{C_4(\{x_i\})-C^\text{Gauss}_4(\{x_i\})}{C_4(\{x_i\})}\,.
    \label{eq:fourpointquantity}
\end{equation}
This is shown for different coupling strengths and inverse temperatures in the 2D density plots of Fig. \ref{fig:fourpoint} obtained by fixing the positions $x_0$ and $x_3$. For high temperatures ($MR\leq1$), we see the effects of quantum fluctuations magnified by the division (in that regime, the full 4-point can be positive or negative, while the product will always remain positive). Yet the largest difference can be observed in the intermediate-temperature regime, around $MR=4$, although the exact regime depends on the coupling strength. At high temperatures, the system is mainly influenced by highly excited states far above the cosine potential, so the interaction terms become negligible. In contrast, at low temperatures, the ground state lies near the potential minimum, where it can be approximated by a parabola; therefore, the correlations are well described by a massive free-field theory. Significant connected correlations emerge only at intermediate temperatures, where thermal states can probe the non-parabolic features of the potential. This can be quantified by introducing a kurtosis-like quantity,  in analogy with  Refs. \cite{2017Natur.545..323S,2018PhRvL.121k0402K}, after integrating over the spatial coordinates:
\begin{equation}
    K(x_0,x_3) = \frac{\int dx_1 \int dx_2\left| C_4(\{x_i\} - C^\text{Gauss}_4(\{x_i\} \right|}{\int dx_1 \int dx_2 \left|C_4(\{x_i\})\right|}\,,
    \label{eq:kurtosis}
\end{equation}
for which we present the numerical results in Fig. \ref{fig:kurtosis}. 
%{\bf JP: Are we integrating over space time or just space coordinates? As written it looks like only space.} \MT{Only spatial coordinates}
We see that the strength of the correlations is largest in the intermediate-temperature regime, and that this regime shifts to higher temperatures (relative to the soliton mass $M$) for larger couplings. Notice also that the strength of the correlations increases with the coupling as well, as expected.

\paragraph{Concluding summary.---}
This work establishes the method of random surfaces (MRS) as an efficient framework for exploring the finite-temperature dynamics of the sine--Gordon model. By moving beyond the calculation of free energies and expectation values, we have demonstrated that the MRS can successfully determine two-point and higher-order correlation functions in regimes where traditional methods often falter.

While existing approaches, such as form-factor expansions, semiclassical methods, %\cite{2013PhRvL.110i0404D,2016PhRvE..93f2101K,2017PhRvL.119j0603M,2019JSMTE..08.4012V,2019PhRvA.100a3613H} 
and truncated Hamiltonian approaches %\cite{2018PhRvL.121k0402K,2019PhRvA.100a3613H,2024PhRvB.109a4308S}, 
are restricted to specific limits, the MRS provides the first reliable results for correlation functions at intermediate values of the coupling and temperature. It would be interesting to compare our approach to a direct Monte Carlo evaluation \cite{1994PhyA..211..255H,Flamino:2018jmo,2025PhRvB.112c5148B,2026PhRvB.113g5126B} of the path integral \eqref{eqn:avg0}.
%\mk{We note that ballistic fluctuation theory \cite{Myers2020,Doyon2019b,DelVecchio2023} based on generalized hydrodynamics cannot access correlators of vertex operators $V_\alpha$ where $\alpha$ is an integer multiple of $\beta$.}

Our numerical results show excellent agreement with known analytical limits. In the high-temperature regime, the correlation length $\xi$ converges to the conformal result {$2R/\beta^2$, 
while at low temperatures, it converges to the inverse mass gap $1/m_1$. }

We also derived an exact formula for arbitrary multipoint functions of vertex operators, provided they satisfy the neutrality condition $\sum_{i}\alpha_{i}=n\beta$. This extension allows for non-perturbative control over the theory's multi-point correlations. Despite the introduction of numerical artifacts like Fourier mode truncation ($m_\text{max}$) and finite system sizes ($L$), the correlation lengths extracted from exponential fits remain remarkably consistent across different geometries and mode cutoffs. We demonstrated that the 4-point function can be used to quantify interaction strength by characterizing the non-Gaussianity of correlations, yielding results that agree with theoretical expectations.

The ability to calculate higher-order functions — such as the 4-point functions presented here — opens the door to a deeper understanding of the sine--Gordon model’s thermal dynamics and its applications in 1+1-dimensional condensed matter systems. While numerical uncertainty persists at very low temperatures due to larger statistical errors, the MRS performs optimally in the intermediate-temperature regime ($MR \approx 1$), offering an efficient approach to investigate previously inaccessible quantum field theory observables. Our method can be used to benchmark quantum simulators of the sine--Gordon model \cite{2021NuPhB.96815445R,PRXQuantum.4.030308,2026JHEP...03..125R}.

\paragraph{Acknowledgments.---} We thank Adilet Imambekov for inspiring this project many years ago and for many insightful discussions. We thank Alvise Bastianello for providing data for the 2-point function in the classical limit. We thank Riccarda Bonsignori for her assistance in preparing a figure.
This work was supported by the National Research, Development and Innovation Office (NKFIH) through the OTKA Grants ANN 142584 and K 138606, and also by the HUN-REN Hungarian Research Network through the Supported Research Groups Programme, HUN-REN-BME-BCE Quantum Technology Research Group (TKCS-2024/34). MT was partially supported by the Doctoral Excellence Fellowship Programme (DCEP), funded by the National Research, Development and Innovation Fund of the Ministry of Culture and Innovation, and the Budapest University
of Technology and Economics, under a grant agreement with the National Research, Development and Innovation Office. GT was also partially supported by the NKFIH grant “Quantum Information National Laboratory” (Grant No. 2022-2.1.1-NL-2022-00004). J.H.P. is partially supported by NSF Grant No. DMR-2515945.

\bibliographystyle{utphys}
\bibliography{sG_corr}

\providecommand{\href}[2]{#2}\begingroup\raggedright\begin{thebibliography}{10}

\bibitem{tsvelik_2003}
A.~M. Tsvelik, \href{https://dx.doi.org/10.1017/CBO9780511615832}{{\em Quantum Field Theory in Condensed Matter Physics}}.
\newblock Cambridge University Press, 2003.

\bibitem{Giamarchi:743140}
T.~Giamarchi, \href{https://dx.doi.org/10.1093/acprof:oso/9780198525004.001.0001}{{\em {Quantum physics in one dimension}}}.
\newblock Clarendon Press, Oxford, 2004.

\bibitem{2007PhRvB..75q4511G}
V.~{Gritsev}, A.~{Polkovnikov}, and E.~{Demler}, ``{Linear response theory for a pair of coupled one-dimensional condensates of interacting atoms},'' \href{https://dx.doi.org/10.1103/PhysRevB.75.174511}{{\em Phys. Rev. B} {\bfseries 75} (2007) 174511}, \href{https://arxiv.org/abs/cond-mat/0701421}{{\ttfamily arXiv:cond-mat/0701421}}.

\bibitem{2010PhRvL.105s0403C}
J.~I. {Cirac}, P.~{Maraner}, and J.~K. {Pachos}, ``{Cold Atom Simulation of Interacting Relativistic Quantum Field Theories},'' \href{https://dx.doi.org/10.1103/PhysRevLett.105.190403}{{\em Phys. Rev. Lett.} {\bfseries 105} (2010) 190403}, \href{https://arxiv.org/abs/1006.2975}{{\ttfamily arXiv:1006.2975}}.

\bibitem{2010Natur.466..597H}
E.~{Haller}, R.~{Hart}, M.~J. {Mark}, J.~G. {Danzl}, L.~{Reichs{\"o}llner}, M.~{Gustavsson}, M.~{Dalmonte}, G.~{Pupillo}, and H.-C. {N{\"a}gerl}, ``{Pinning quantum phase transition for a Luttinger liquid of strongly interacting bosons},'' \href{https://dx.doi.org/10.1038/nature09259}{{\em Nature} {\bfseries 466} (2010) 597--600}, \href{https://arxiv.org/abs/1004.3168}{{\ttfamily arXiv:1004.3168}}.

\bibitem{2017Natur.545..323S}
T.~{Schweigler}, V.~{Kasper}, S.~{Erne}, I.~{Mazets}, B.~{Rauer}, F.~{Cataldini}, T.~{Langen}, T.~{Gasenzer}, J.~{Berges}, and J.~{Schmiedmayer}, ``{Experimental characterization of a quantum many-body system via higher-order correlations},'' \href{https://dx.doi.org/10.1038/nature22310}{{\em Nature} {\bfseries 545} (2017) 323--326}, \href{https://arxiv.org/abs/1505.03126}{{\ttfamily arXiv:1505.03126}}.

\bibitem{2024PhRvB.109c5118B}
A.~{Bastianello}, ``{Sine-Gordon model from coupled condensates: A generalized hydrodynamics viewpoint},'' \href{https://dx.doi.org/10.1103/PhysRevB.109.035118}{{\em \prb} {\bfseries 109} (Jan., 2024) 035118}, \href{https://arxiv.org/abs/2310.04493}{{\ttfamily arXiv:2310.04493}}.

\bibitem{Controzzi2001}
D.~Controzzi, F.~H.~L. Essler, and A.~M. Tsvelik, \href{https://dx.doi.org/10.1007/978-94-010-0838-9_2}{``{Dynamical Properties of One Dimensional Mott Insulators},''} in {\em New Theoretical Approaches to Strongly Correlated Systems}, A.~M. Tsvelik, ed., p.~25–46.
\newblock Springer Netherlands, Dordrecht, 2001.
\newblock \href{https://arxiv.org/abs/cond-mat/0011439}{{\ttfamily arXiv:cond-mat/0011439}}.

\bibitem{2005ffsc.book..684E}
F.~H.~L. {Essler} and R.~M. {Konik}, \href{https://dx.doi.org/10.1142/9789812775344_0020}{``{Application of Massive Integrable Quantum Field Theories to Problems in Condensed Matter Physics},''} in {\em From Fields to Strings: Circumnavigating Theoretical Physics: Ian Kogan Memorial Collection (in 3 Vols)}, {M. Shifman et al.}, ed., pp.~684--830.
\newblock {World Scientific}, 2005.
\newblock \href{https://arxiv.org/abs/cond-mat/0412421}{{\ttfamily arXiv:cond-mat/0412421}}.

\bibitem{2021NuPhB.96815445R}
A.~{Roy}, D.~{Schuricht}, J.~{Hauschild}, F.~{Pollmann}, and H.~{Saleur}, ``{The quantum sine-Gordon model with quantum circuits},'' \href{https://dx.doi.org/10.1016/j.nuclphysb.2021.115445}{{\em Nucl. Phys. B} {\bfseries 968} (2021) 115445}, \href{https://arxiv.org/abs/2007.06874}{{\ttfamily arXiv:2007.06874}}.

\bibitem{Wybo2022}
E.~Wybo, M.~Knap, and A.~Bastianello, ``{Quantum sine-Gordon dynamics in coupled spin chains},'' \href{https://dx.doi.org/10.1103/PhysRevB.106.075102}{{\em Phys. Rev. B} {\bfseries 106} (2022) 075102}, \href{https://arxiv.org/abs/2203.09530}{{\ttfamily arXiv:2203.09530}}.

\bibitem{PRXQuantum.4.030308}
E.~Wybo, A.~Bastianello, M.~Aidelsburger, I.~Bloch, and M.~Knap, ``{Preparing and Analyzing Solitons in the Sine-Gordon Model with Quantum Gas Microscopes},'' \href{https://dx.doi.org/10.1103/PRXQuantum.4.030308}{{\em PRX Quantum} {\bfseries 4} (2023) 030308}, \href{https://arxiv.org/abs/2303.16221}{{\ttfamily arXiv:2303.16221}}.

\bibitem{1975PhRvD..11.2088C}
S.~{Coleman}, ``{Quantum sine-Gordon equation as the massive Thirring model},'' \href{https://dx.doi.org/10.1103/PhysRevD.11.2088}{{\em Phys. Rev. D} {\bfseries 11} (1975) 2088--2097}.

\bibitem{1975PhRvD..11.3026M}
S.~{Mandelstam}, ``{Soliton operators for the quantized sine-Gordon equation},'' \href{https://dx.doi.org/10.1103/PhysRevD.11.3026}{{\em Phys. Rev. D} {\bfseries 11} (1975) 3026--3030}.

\bibitem{1977CMaPh..55..183Z}
A.~B. {Zamolodchikov}, ``{Exact two-particle S-matrix of quantum sine-Gordon solitons},'' \href{https://dx.doi.org/10.1007/BF01626520}{{\em Commun. Math. Phys.} {\bfseries 55} (1977) 183--186}.

\bibitem{ZAMOLODCHIKOV1979253}
A.~B. Zamolodchikov and A.~B. Zamolodchikov, ``{Factorized S-matrices in two dimensions as the exact solutions of certain relativistic quantum field theory models},'' \href{https://dx.doi.org/https://doi.org/10.1016/0003-4916(79)90391-9}{{\em Ann. Phys.} {\bfseries 120} (1979) 253--291}.

\bibitem{1991JPhA...24.3111K}
A.~{Kl\"umper}, M.~T. {Batchelor}, and P.~A. {Pearce}, ``{Central charges of the 6- and 19-vertex models with twisted boundary conditions},'' \href{https://dx.doi.org/10.1088/0305-4470/24/13/025}{{\em J. Phys. A Math. Gen.} {\bfseries 24} (1991) 3111--3133}.

\bibitem{1995NuPhB.438..413D}
C.~{Destri} and H.~J. {de Vega}, ``{Unified approach to Thermodynamic Bethe Ansatz and finite size corrections for lattice models and field theories},'' \href{https://dx.doi.org/10.1016/0550-3213(94)00547-R}{{\em Nucl. Phys. B} {\bfseries 438} (1995) 413--454}, \href{https://arxiv.org/abs/hep-th/9407117}{{\ttfamily arXiv:hep-th/9407117}}.

\bibitem{Hegedus2025}
{\'A}.~{Heged{\H{u}}s}, ``{Thermodynamics in the sine-Gordon model: the NLIE approach},'' \href{https://dx.doi.org/10.1016/j.nuclphysb.2025.117155}{{\em Nuclear Physics B} {\bfseries 1020} (2025) 117155}, \href{https://arxiv.org/abs/2507.19200}{{\ttfamily arXiv:2507.19200}}.

\bibitem{Hegedus2026}
{\'A}.~{Heged{\H{u}}s}, ``{NLIE formulations for the generalized Gibbs ensemble in the sine-Gordon model},'' \href{https://dx.doi.org/10.1016/j.nuclphysb.2026.117385}{{\em Nuclear Physics B} {\bfseries 1025} (2026) 117385}, \href{https://arxiv.org/abs/2510.25344}{{\ttfamily arXiv:2510.25344}}.

\bibitem{koch2023exact}
R.~Koch and A.~Bastianello, ``{Exact Thermodynamics and Transport in the Classical Sine-Gordon Model},'' \href{https://dx.doi.org/10.21468/SciPostPhys.15.4.140}{{\em SciPost Phys.} {\bfseries 15} (2023) 140}, \href{https://arxiv.org/abs/2303.16932}{{\ttfamily arXiv:2303.16932}}.

\bibitem{2024ScPP...16..145N}
B.~C. {Nagy}, G.~{Tak{\'a}cs}, and M.~{Kormos}, ``{Thermodynamic Bethe Ansatz and generalised hydrodynamics in the sine-Gordon model},'' \href{https://dx.doi.org/10.21468/SciPostPhys.16.6.145}{{\em SciPost Phys.} {\bfseries 16} (2024) 145}, \href{https://arxiv.org/abs/2312.03909}{{\ttfamily arXiv:2312.03909}}.

\bibitem{1992ASMP...14....1S}
F.~A. {Smirnov}, ``{Completely Integrable Models of Quantum Field Theory},'' \href{https://dx.doi.org/10.1142/9789812798312_0001}{{\em Form Factors In Completely Integrable Models Of Quantum Field Theory. Series: Advanced Series in Mathematical Physics} {\bfseries 14} (1992) 1--5}.

\bibitem{Lukyanov1996}
S.~{Lukyanov} and A.~{Zamolodchikov}, ``{Exact expectation values of local fields in the quantum sine-Gordon model},'' \href{https://dx.doi.org/10.1016/S0550-3213(97)00123-5}{{\em Nucl. Phys. B} {\bfseries 493} (1997) 571--587}, \href{https://arxiv.org/abs/hep-th/9611238}{{\ttfamily arXiv:hep-th/9611238}}.

\bibitem{2014JHEP...03..026B}
F.~{Buccheri} and G.~{Tak{\'a}cs}, ``{Finite temperature one-point functions in non-diagonal integrable field theories: the sine-Gordon model},'' \href{https://dx.doi.org/10.1007/JHEP03(2014)026}{{\em J. High Energ. Physics} {\bfseries 2014} (2014) 26}, \href{https://arxiv.org/abs/1312.2623}{{\ttfamily arXiv:1312.2623}}.

\bibitem{2009JSMTE..09..018E}
F.~H.~L. {Essler} and R.~M. {Konik}, ``{Finite-temperature dynamical correlations in massive integrable quantum field theories},'' \href{https://dx.doi.org/10.1088/1742-5468/2009/09/P09018}{{\em J. Stat. Mech.} {\bfseries 2009} (2009) 09018}, \href{https://arxiv.org/abs/0907.0779}{{\ttfamily arXiv:0907.0779}}.

\bibitem{2010JSMTE..11..012P}
B.~{Pozsgay} and G.~{Tak{\'a}cs}, ``{Form factor expansion for thermal correlators},'' \href{https://dx.doi.org/10.1088/1742-5468/2010/11/P11012}{{\em J. Stat. Mech.} {\bfseries 2010} (2010) 11012}, \href{https://arxiv.org/abs/1008.3810}{{\ttfamily arXiv:1008.3810}}.

\bibitem{Yurov:1989yu}
V.~P. Yurov and A.~B. Zamolodchikov, ``{Truncated Conformal Space Approach to Scaling Lee-Yang Model},'' \href{https://dx.doi.org/10.1142/S0217751X9000218X}{{\em Int. J. Mod. Phys. A} {\bfseries 5} (1990) 3221--3246}.

\bibitem{1998PhLB..430..264F}
G.~{Feverati}, F.~{Ravanini}, and G.~{Tak{\'a}cs}, ``{Truncated conformal space at c=1, nonlinear integral equation and quantization rules for multi-soliton states},'' \href{https://dx.doi.org/10.1016/S0370-2693(98)00543-7}{{\em Physics Letters B} {\bfseries 430} (1998) 264--273}, \href{https://arxiv.org/abs/hep-th/9803104}{{\ttfamily arXiv:hep-th/9803104}}.

\bibitem{2018PhRvL.121k0402K}
I.~{Kukuljan}, S.~{Sotiriadis}, and G.~{Tak\'acs}, ``{Correlation Functions of the Quantum Sine-Gordon Model in and out of Equilibrium},'' \href{https://dx.doi.org/10.1103/PhysRevLett.121.110402}{{\em Phys. Rev. Lett.} {\bfseries 121} (2018) 110402}, \href{https://arxiv.org/abs/1802.08696}{{\ttfamily arXiv:1802.08696}}.

\bibitem{2018RPPh...81d6002J}
A.~J.~A. {James}, R.~M. {Konik}, P.~{Lecheminant}, N.~J. {Robinson}, and A.~M. {Tsvelik}, ``{Non-perturbative methodologies for low-dimensional strongly-correlated systems: From non-Abelian bosonization to truncated spectrum methods},'' \href{https://dx.doi.org/10.1088/1361-6633/aa91ea}{{\em Reports on Progress in Physics} {\bfseries 81} (2018) 046002}, \href{https://arxiv.org/abs/1703.08421}{{\ttfamily arXiv:1703.08421}}.

\bibitem{2022CoPhC.27708376H}
D.~X. {Horv{\'a}th}, K.~{H{\'o}ds{\'a}gi}, and G.~{Tak{\'a}cs}, ``{Chirally factorised truncated conformal space approach},'' \href{https://dx.doi.org/10.1016/j.cpc.2022.108376}{{\em Computer Physics Communications} {\bfseries 277} (2022) 108376}, \href{https://arxiv.org/abs/2201.06509}{{\ttfamily arXiv:2201.06509}}.

\bibitem{2020arXiv200513544A}
N.~{Anand}, A.~L. {Fitzpatrick}, E.~{Katz}, Z.~U. {Khandker}, M.~T. {Walters}, and Y.~{Xin}, ``{Introduction to Lightcone Conformal Truncation: QFT Dynamics from CFT Data},'' \href{https://dx.doi.org/10.48550/arXiv.2005.13544}{{\em arXiv e-prints} (2020) }, \href{https://arxiv.org/abs/2005.13544}{{\ttfamily arXiv:2005.13544}}.

\bibitem{2005PhRvL..95r7201D}
K.~{Damle} and S.~{Sachdev}, ``{Universal Relaxational Dynamics of Gapped One-Dimensional Models in the Quantum Sine-Gordon Universality Class},'' \href{https://dx.doi.org/10.1103/PhysRevLett.95.187201}{{\em Phys. Rev. Lett.} {\bfseries 95} (2005) 187201}, \href{https://arxiv.org/abs/cond-mat/0507380}{{\ttfamily arXiv:cond-mat/0507380}}.

\bibitem{2017PhRvL.119j0603M}
C.~P. {Moca}, M.~{Kormos}, and G.~{Zar{\'a}nd}, ``{Hybrid Semiclassical Theory of Quantum Quenches in One-Dimensional Systems},'' \href{https://dx.doi.org/10.1103/PhysRevLett.119.100603}{{\em Phys. Rev. Lett.} {\bfseries 119} (2017) 100603}, \href{https://arxiv.org/abs/1609.00974}{{\ttfamily arXiv:1609.00974}}.

\bibitem{PhysRevB.106.205151}
M.~{Kormos}, D.~{V{\"o}r{\"o}s}, and G.~{Zar{\'a}nd}, ``{Finite-temperature dynamics in gapped one-dimensional models in the sine-Gordon family},'' \href{https://dx.doi.org/10.1103/PhysRevB.106.205151}{{\em \prb} {\bfseries 106} (2022) 205151}, \href{https://arxiv.org/abs/2208.08406}{{\ttfamily arXiv:2208.08406}}.

\bibitem{Myers2020}
J.~Myers, J.~Bhaseen, R.~J. Harris, and B.~Doyon, ``{Transport fluctuations in integrable models out of equilibrium},'' \href{https://dx.doi.org/10.21468/SciPostPhys.8.1.007}{{\em SciPost Phys.} {\bfseries 8} (2020) 007}, \href{https://arxiv.org/abs/1812.02082}{{\ttfamily arXiv:1812.02082}}.

\bibitem{Doyon2019b}
B.~Doyon and J.~Myers, ``{Fluctuations in Ballistic Transport from Euler Hydrodynamics},'' \href{https://dx.doi.org/10.1007/s00023-019-00860-w}{{\em Ann. Henri Poincar{\'{e}}} {\bfseries 21} (2020) 255--302}, \href{https://arxiv.org/abs/1902.00320}{{\ttfamily arXiv:1902.00320}}.

\bibitem{DelVecchio2023}
G.~D.~V. {Del Vecchio}, M.~{Kormos}, B.~{Doyon}, and A.~{Bastianello}, ``{Exact Large-Scale Fluctuations of the Phase Field in the Sine-Gordon Model},'' \href{https://dx.doi.org/10.1103/PhysRevLett.131.263401}{{\em \prl} {\bfseries 131} (2023) 263401}, \href{https://arxiv.org/abs/2305.10495}{{\ttfamily arXiv:2305.10495}}.

\bibitem{Toth2025}
M.~{T{\'o}th}, J.~H. {Pixley}, D.~{Sz{\'a}sz-Schagrin}, G.~{Tak{\'a}cs}, and M.~{Kormos}, ``{Sine-Gordon model at finite temperature: The method of random surfaces},'' \href{https://dx.doi.org/10.1103/PhysRevB.111.155112}{{\em Phys. Rev. B} {\bfseries 111} (2025) 155112}, \href{https://arxiv.org/abs/2408.08828}{{\ttfamily arXiv:2408.08828}}.

\bibitem{2008PhRvA..77f3606I}
A.~{Imambekov}, V.~{Gritsev}, and E.~{Demler}, ``{Mapping of Coulomb gases and sine-Gordon models to statistics of random surfaces},'' \href{https://dx.doi.org/10.1103/PhysRevA.77.063606}{{\em Phys. Rev. A} {\bfseries 77} (2008) 063606}, \href{https://arxiv.org/abs/cond-mat/0612011}{{\ttfamily arXiv:cond-mat/0612011}}.

\bibitem{2008NatPh...4..489H}
S.~{Hofferberth}, I.~{Lesanovsky}, T.~{Schumm}, A.~{Imambekov}, V.~{Gritsev}, E.~{Demler}, and J.~{Schmiedmayer}, ``{Probing quantum and thermal noise in an interacting many-body system},'' \href{https://dx.doi.org/10.1038/nphys941}{{\em Nat. Phys.} {\bfseries 4} (2008) 489--495}, \href{https://arxiv.org/abs/0710.1575}{{\ttfamily arXiv:0710.1575}}.

\bibitem{2010PhRvA..82c2118O}
P.~P. {Orth}, A.~{Imambekov}, and K.~{Le Hur}, ``{Universality in dissipative Landau-Zener transitions},'' \href{https://dx.doi.org/10.1103/PhysRevA.82.032118}{{\em Phys. Rev. A} {\bfseries 82} (2010) 032118}, \href{https://arxiv.org/abs/0912.3531}{{\ttfamily arXiv:0912.3531}}.

\bibitem{2013PhRvB..87a4305O}
P.~P. {Orth}, A.~{Imambekov}, and K.~{Le Hur}, ``{Nonperturbative stochastic method for driven spin-boson model},'' \href{https://dx.doi.org/10.1103/PhysRevB.87.014305}{{\em Phys. Rev. B} {\bfseries 87} (2013) 014305}, \href{https://arxiv.org/abs/1211.1201}{{\ttfamily arXiv:1211.1201}}.

\bibitem{SM}
Supplemental Material containing (1) the derivation of the two-point function using functional derivatives, and (2) details of the numerical implementation.

\bibitem{1995IJMPA..10.1125Z}
A.~B. {Zamolodchikov}, ``{Mass Scale in the Sine-Gordon Model and its Reductions},'' \href{https://dx.doi.org/10.1142/S0217751X9500053X}{{\em Int. J. Mod. Phys. A} {\bfseries 10} (1995) 1125--1150}.

\bibitem{1994PhyA..211..255H}
M.~{Hasenbusch}, M.~{Marcu}, and K.~{Pinn}, ``{The sine Gordon model: perturbation theory and cluster Monte Carlo},'' \href{https://dx.doi.org/10.1016/0378-4371(94)00196-0}{{\em Physica A Stat. Mech. Appl.} {\bfseries 211} (1994) 255--278}, \href{https://arxiv.org/abs/hep-lat/9408005}{{\ttfamily arXiv:hep-lat/9408005}}.

\bibitem{Flamino:2018jmo}
J.~Flamino and J.~Giedt, ``{Lattice sine-Gordon model},'' \href{https://dx.doi.org/10.1103/PhysRevD.101.074503}{{\em Phys. Rev. D} {\bfseries 101} (2020) 074503}, \href{https://arxiv.org/abs/1811.01219}{{\ttfamily arXiv:1811.01219}}.

\bibitem{2025PhRvB.112c5148B}
O.~{Bouverot-Dupuis}, A.~{Rosso}, and M.~{Michel}, ``{Bosonized one-dimensional quantum systems through enhanced event-chain Monte Carlo},'' \href{https://dx.doi.org/10.1103/k43n-yz82}{{\em \prb} {\bfseries 112} (2025) 035148}, \href{https://arxiv.org/abs/2503.11577}{{\ttfamily arXiv:2503.11577}}.

\bibitem{2026PhRvB.113g5126B}
O.~{Bouverot-Dupuis}, L.~{Foini}, and A.~{Rosso}, ``{Generic Mott transition in the sine-Gordon model through an embedded worm algorithm},'' \href{https://dx.doi.org/10.1103/4k6t-lnzp}{{\em \prb} {\bfseries 113} (2026) 075126}, \href{https://arxiv.org/abs/2510.20901}{{\ttfamily arXiv:2510.20901}}.

\bibitem{2026JHEP...03..125R}
T.~{Rainaldi}, V.~{Ale}, M.~{Grau}, D.~{Kharzeev}, E.~{Rico}, F.~{Ringer}, P.~{Shome}, and G.~{Siopsis}, ``{Trigonometric continuous-variable gates and hybrid quantum simulations of the sine-Gordon model},'' \href{https://dx.doi.org/10.1007/JHEP03(2026)125}{{\em J. High Energ. Phys.} {\bfseries 2026} (2026) 125}, \href{https://arxiv.org/abs/2512.19582}{{\ttfamily arXiv:2512.19582}}.

\end{thebibliography}\endgroup

\clearpage

\appendix

\onecolumngrid

\centerline{\large\emph{End Matter}}

\section{Exact calculation of multi-point functions}
To establish a general framework for non-perturbative observables, we extend the MRS framework to calculate general multi-point functions of the form
\begin{align}
    C_{\{\alpha_i\}}(\br_1, ...,\br_N) = \expv{ \prod_i \hat V_{\alpha_i}(\br_i)} = {Z^{-1}Z_0}\expv{\prod_i a^{-\frac{\alpha_i^2}{4\pi}} e^{i\alpha_i\hat\phi(\br_i)}e^{-S_I(\lambda)}}_0,
\end{align}
where $Z$ is the partition function, $Z_I$ is the interacting part of the partition function defined in \eqref{eq:genfunc}, and 
\begin{equation}
    \hat V_{\alpha_i}(\br_i) = a^{-\frac{\alpha_i^2}{4\pi}} e^{i\alpha_i\hat\phi(\br_i)}
\end{equation}
are the vertex operators with $a$ being a short-distance regulator. Expanding the exponential interaction term $e^{-S_I}$ in powers of the bare coupling constant $\lambda_0$ yields:
\begin{multline}
   C_{\{\alpha_i\}}(\br_1, ...,\br_N)  = \frac{{1}}{Z_I(\lambda)} \sum_{n=0}^\infty \frac{1}{n!}\left(\frac{\lambda_0}2\right)^n \prod_{j=1}^n \int dx_j \int_0^R d\tau_j
   \prod_{i=1}^N a^{-\frac{\alpha_i^2}{4\pi}}\,\Expv{\prod_{i=1}^N e^{i\alpha_i \phi(\br_i)} \prod_{j=1}^n \left( e^{i\beta\phi(\boldsymbol{\rho}_j)} + e^{-i\beta\phi(\boldsymbol{\rho}_j)} \right)}_0\,.
\end{multline}
with $\boldsymbol{\rho}_j = (x_j, \tau_j)$. The free expectation value of products of exponentials is nonzero if the neutrality condition is satisfied, i.e. the some of exponents is zero. This can happen if $\sum_{i}\alpha_{i}=s\beta$ for some $s \in \mathbb{Z}$. For a given expansion order $n$, this requires $n_{+}=(n-s)/2$ positive and $n_{-}=(n+s)/2$ negative exponentials from the interaction term. The resulting free-field expectation value of these exponentials generates a sum of Green's functions $G(\br_k, \br_l)$ in the exponent:
\begin{align}
    C_{\{\alpha_i\}}(\br_1, ...,\br_N)  &= \frac{{1}}{Z_I(\lambda)} \sideset{}{'}\sum_{n=0}^\infty \frac{1}{n_+!\,n_-!} \left(\frac{\lambda}{2R^\Delta}\right)^n C_{n, \{\alpha_i\}}(\br_1, ...,\br_N),
    \label{eq:npoint}
\end{align}
where
\begin{align}
    C_{n, \{\alpha_i\}}(\br_1, ...,\br_N) &= \prod_{j=1}^n \int d\boldsymbol{\rho}_j 
    \,\prod_{i=1}^N a^{-\frac{\alpha_i^2}{4\pi}}\,
    \bigg\langle \prod_{i=1}^N e^{is_i\beta\phi(\br_i)} 
    \prod_{j=1}^{n_+}e^{i\beta\phi(\boldsymbol{\rho}_j)} 
    \prod_{j=1}^{n_-}e^{-i\beta\phi(\boldsymbol{\rho}_j)} \bigg\rangle_0 
    \nonumber  \\ & = 
    R^{-\Delta_\Sigma}\prod_{j=1}^n \int d\boldsymbol{\rho}_j \,
    e^{-\Delta \sum_{k<l}^n \epsilon_k \epsilon_l G(\boldsymbol{\rho}_k, \boldsymbol{\rho}_l)} e^{-\Delta\sum_i^N   \sum_k^n  s_i\epsilon_k G(\br_i,\boldsymbol{\rho}_k)} e^{-\Delta\sum_{i<j}^N s_i s_j  G(\br_i,\br_j)},
\end{align}
where $\Delta_\Sigma = \sum_i \frac{\alpha_i^2}{4\pi}$, $s_i=\alpha_i/\beta$, and $\lambda=\lambda_0 a^{\beta^2/4\pi}$.

We decouple these terms by first employing a 2D Fourier expansion of the Green's function 
\begin{equation}
    G(\br_1, \br_2) = \sum_{\bff} G_{\bff} \psi_{\bff}(\br_1) \psi_{\bff}(\br_2)\,,
\end{equation} 
and then performing a Hubbard--Stratonovich (HS) decoupling for each mode using the identity
\begin{align}
    e^{-\frac{1}{2}x^2} = \frac1{\sqrt{2\pi}} \int_{-\infty}^\infty e^{-\frac{y^2}{2} \pm ixy} dy\,.
\end{align}
This procedure leads to the factorization of the multidimensional space-time integral at the price of introducing an integral of a HS variable $t_\ff$ for each mode. 
%auxiliary Gaussian random variables $\{t_\ff\}$
%effectively linearizing the quadratic dependence on the field coordinates. 
This leads to
\begin{align}
    C_{n, \{\alpha_i\}}(\br_1, ...,\br_N) = R^{-\Delta_\Sigma} \int \mathcal{D}[t_\ff] \,g(\{t_\ff\})^{n_+} g(\{-t_\ff\})^{n_-}  \prod_{i=1}^N
     h_{\alpha_i}(\{t_\ff\}, \br_i)\,,
\end{align}
where $\int \mathcal{D}[t_\ff] = \prod_\ff \int_{-\infty}^\infty \frac{dt_\ff}{\sqrt{2\pi}} e^{-t_\ff^2/2}$, and
$g(\{t_\ff\}) = \int d\br \, h(\{t_\ff\},\br)$ with 
\begin{align}
    h_{\alpha_i}(\{t_\ff\}, \br_i) &= C_{s_i} \exp(\sum_\ff\left[ i t_\ff s_i\sqrt{\Delta G_\ff} \psi_\ff(\br_i) \right])\,\\
    C_{s_i}&=\exp{\Delta s_i^2/2\,\sum_{m,n}A_{mn}}=C^{s_i^2}
    \end{align}
describing the local coupling of the $i$-th vertex operator to the random surface modes $\{t_\ff\}$. 

Substituting everything into \eqref{eq:npoint}, the perturbation series over $n$ can be summed up, which leads to the master formula for the $N$-point correlation function:
\begin{align}
     \expv{ \prod_i \hat V_{\alpha_i}(\br_i)} = \frac{R^{-\Delta_\Sigma}}{Z_I(\lambda)} \int \mathcal{D}[t_\ff]\, 
    \left(\frac{g(\{-t_\ff\})}{g(\{t_\ff\})}\right)^{\frac s{2}} \prod_i
    h_{\alpha_i}(\{t_\ff\}, \br_i)\,
    I_{\left|s\right|}\left( \lambda R^{-\Delta} \sqrt{g(\{t_\ff\})g(\{-t_\ff\})}\right)\,,
    \label{eq:npointfinal}
\end{align}
where $I_{|s|}$ is the modified Bessel function of the first kind, whose order $s = \sum_{i}\alpha_{i}/\beta$ is determined by the exponents of the vertex operators. 
%with $s = \sum_i \alpha_i/\beta$, and $\Delta_{\Sigma}=\sum_{i}\frac{\alpha_{i}^{2}}{4\pi}$ accounts for the collective scaling dimensions of the operators.

%This result shows that the modified Bessel function $I_{|s|}$ serves as the weighting kernel for the sector $s$. 
This general expression reduces to the standard MRS expectation value when $N=1$, while for the special case of $N=2$, $\alpha_1=-\alpha_2=\beta$, it yields the two-point neutral correlator discussed in the main text. Furthermore, it provides a direct computational approach for higher-order correlations, such as the 4-point function shown in Fig. \ref{fig:fourpoint}.

\section{Calculation of $K_4$}
To evaluate the 4-point vertex operator correlation function, the free-field expectation value serves as a baseline. Starting from the free expectation value of a product of vertex operators evaluated using the ``mulitplicative Wick's theorem''
\begin{equation}
    \expv{\prod_{j=0}^3 e^{i\alpha_j\hat\phi(x_j)}}_0 = e^{-\sum_{j<k}^3\alpha_j\alpha_k\tilde{G}(x_j,x_k)-\sum_{j=0}^3\alpha_j^2\tilde{G}(x_j,x_j)}\,,
\end{equation}
and substituting the Green's function on an infinite cylinder gives
\begin{equation}
    \expv{\prod_{j=0}^3 e^{i\alpha_j\hat\phi(x_j)}}_0 = a^{\frac{\sum_{j=0}^3\alpha_j^2}{4\pi}} \left(\frac{1}{R}\right)^{\frac{\left(\sum_{j=0}^3\alpha_j\right)^2}{4\pi}} \prod_{j<k} \left|\frac{R}{\pi}\sinh\left(\frac{\pi}{R}(x_k-x_j+a)\right)\right|^{\frac{\alpha_j\alpha_k}{2\pi}}\,.
\end{equation}
When recasting this as a product of two-point functions of vertex operators, the precise representation must be chosen carefully to avoid incorrect scaling contributions regarding system size $R$ and the UV cut-off $a$. In our case of interest, where $\{\alpha_j\} = \{\beta,\beta,-\beta,-\beta\}$, the problematic factors turn out to be the non-neutral two-point functions $\expv{e^{i\beta\hat\phi(x_0)}e^{i\beta\hat\phi(x_1)}}_0$ and $\expv{e^{-i\beta\hat\phi(x_2)}e^{-i\beta\hat\phi(x_3)}}_0$. Notably, the space-time dependence of these two-point functions matches the reciprocal of the neutral two-point correlators, specifically $\langle e^{i\beta\hat{\phi}(x_{0})}e^{-i\beta\hat{\phi}(x_{1})}\rangle_{0}^{-1}$ and $\langle e^{i\beta\hat{\phi}(x_{2})}e^{-i\beta\hat{\phi}(x_{3})}\rangle_{0}^{-1}$. Furthermore, the prefactors associated with these reciprocal terms effectively cancel the extraneous $a$ and $R$ contributions arising from the product of the neutral two-point functions. The resulting decomposition 
\begin{equation}
    \expv{\prod_{j=0}^3 e^{i\alpha_j\hat\phi(x_j)}}_0 = \frac{\expv{e^{i\beta\hat\phi(x_0)}e^{-i\beta\hat\phi(x_2)}}_0\expv{e^{i\beta\hat\phi(x_0)}e^{-i\beta\hat\phi(x_3)}}_0\expv{e^{i\beta\hat\phi(x_1)}e^{-i\beta\hat\phi(x_2)}}_0\expv{e^{i\beta\hat\phi(x_1)}e^{-i\beta\hat\phi(x_3)}}_0}{\expv{e^{i\beta\hat\phi(x_0)}e^{-i\beta\hat\phi(x_1)}}_0\expv{e^{i\beta\hat\phi(x_2)}e^{-i\beta\hat\phi(x_3)}}_0}
    \label{eq:}
\end{equation}
can be used to construct the free-field reference (disconnected part) for the full interacting 4-point function. We close by noting that while our present considerations were illustrated using the massless limit, they are true for exponentials of any free scalar field (i.e., those with Gaussian correlations), as they rely solely on the validity of Wick's theorem.

\clearpage

\begin{center}
    \Large\emph{Supplemental material}
\end{center}

\section{Correlation function using functional derivatives}

The $1+1$-dimensional sine--Gordon model at finite temperature can be described by the Euclidean action:
\begin{equation}
    S = \int d^2x \left[\frac{1}{2}(\nabla \phi)^2 - \lambda_0\left(\frac{\sigma_1(x)}{2}e^{i \beta \phi} + \frac{\sigma_2(x)}{2}e^{-i \beta \phi}\right)\right] = S_0 + S_I\,,
\end{equation}
where we have introduced space-dependent couplings $\sigma_1(x)$ and $\sigma_2(x)$ to facilitate the functional derivative approach. The interacting part of the partition function, $Z_I$, is defined relative to the free bosonic part $Z_0$:
\begin{equation}
    Z_I = \frac{\int \mathcal{D}\phi e^{-S_0-S_I}}{\int \mathcal{D}\phi e^{-S_0}} = \int \mathcal{D}[t_\ff] I_0\left(\lambda R^{-\Delta}\sqrt{g_1(\{t_\ff\})g_2(\{-t_\ff\})}\right)\,,
\label{eq:genfunc}
\end{equation}
where
\begin{equation}
    g_j(\{t_\ff\}) = C\int d\br\, \sigma_j(\br)\exp\left\{ \sum_\ff^{m_{\mathrm{max}}}  it_\ff\sqrt{\Delta G_\ff}\,\psi_\ff(\br) 
    \right\}
\end{equation}
contains the spatial dependence of the couplings, and $C$ is a constant prefactor. 

Vertex operator expectation values and correlation functions are obtained by differentiating $Z_I$ with respect to these local couplings. The one-point function (expectation value) is given by: 
\begin{equation}
\langle e^{i\beta\phi(x_1)} \rangle = 2 \frac{\delta Z_I}{\delta \sigma_1(x_1)} = \int \mathcal{D}[t_\ff] \sqrt{\frac{g_2(\{-t_\ff\})}{g_1(\{t_\ff\})}} h(\{t_\ff\}, x_1) I_1\left(\lambda\sqrt{g_1(\{t_\ff\})g_2(\{-t_\ff\})}\right)\,,
\end{equation}
where
\begin{equation}
    h(\{t_\ff\}, \br)=\lambda \frac{\delta g_j(\{t_\ff\})}{\delta \sigma_j(\br)} = C \exp(\sum_\ff\left[ i t_\ff \sqrt{\Delta G_\ff} \psi_\ff(\br) \right])\,.
\end{equation}
Evaluating this at the physical limit $\sigma_1 = \sigma_2 = 1$ yields the standard MRS result \cite{Toth2025} for the expectation value:
\begin{equation}
\langle e^{i\beta\phi(x_1)}\rangle = \int D[t_\ff]\sqrt{\frac{g(\{-t_\ff\})}{g(\{t_\ff\})}} h(\{t_\ff\}, x_1) I_1\left(\lambda \sqrt{g_(\{t_\ff\}) g_(\{-t_\ff\})}\right).
\end{equation}
For the two-point function, we perform a second differentiation with respect to $\sigma_2(x_2)$:
\begin{equation}
\langle e^{i\beta\phi(x_1)} e^{-i\beta\phi(x_2)} \rangle = 4 \frac{\delta^2 Z_I}{\delta \sigma_1(x_1) \delta \sigma_2(x_2)}\,.
\end{equation}
Using the Bessel function identity $\frac{dI_\nu(z)}{dz} = I_{\nu-1}(z) - \frac{\nu}{z}I_\nu(z)$ for $\nu=1$, the above expression simplifies to:
\begin{equation}\langle \hat{V}_\beta(x_1) \hat{V}_{-\beta}(x_2) \rangle = \int \mathcal{D}[t_\ff] I_0\left(\lambda\sqrt{g(\{t_\ff\})g(\{-t_\ff\})}\right) h(\{t_\ff\}, x_1) h(\{-t_\ff\}, x_2)\,.
\end{equation}
This result is identical to the neutral correlator derived via the perturbative expansion, confirming its consistency with the functional derivative approach.

\section{Numerical implementation of the MRS}

Here, we provide a summary of the numerical implementation and the systematic treatment of artifacts inherent in the Method of Random Surfaces (MRS).

\begin{figure*}[h]
    \centering
    \begin{subfigure}[b]{0.45\linewidth}
        \centering
        \includegraphics[width=\linewidth]{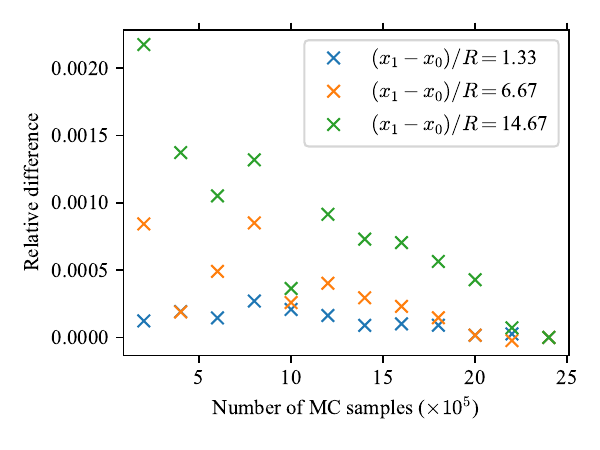}
    \end{subfigure}
    \begin{subfigure}[b]{0.45\linewidth}
        \centering
        \includegraphics[width=\linewidth]{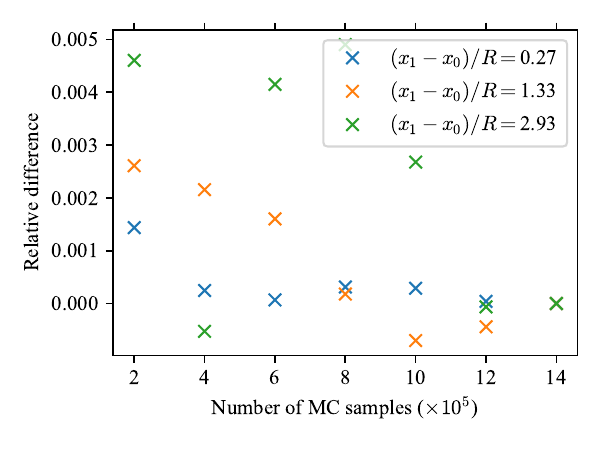}
    \end{subfigure}
    \caption{Convergence of the Monte Carlo (MC) integration as a function of the number of samples. The relative difference compared to the final converged result is shown for weak coupling ($\Delta=0.008$, left) and intermediate coupling ($\Delta=0.08$, right) at three distinct spatial separations. All data points were obtained at a fixed dimensionless inverse temperature of $MR=3$. The results demonstrate that the relative error increases with both the coupling strength and the spatial separation, requiring a higher density of MC samples (reaching a few times $10^6$) to maintain statistical precision in the stronger coupling regimes.
    }
    \label{fig:MCdependence}
\end{figure*}

The fundamental computational step involves drawing a set of random variables $\{t_\ff\}$ from a Gaussian distribution with zero mean and unit variance. These variables correspond to the Fourier modes of the random surface. The primary results were obtained using approximately $1.5 \times 10^{6}$ Monte Carlo (MC) samples, ensuring statistical errors remain smaller than the symbol sizes in the plots. The dependence on the number of MC samples is investigated in Fig.\ \ref{fig:MCdependence}. Convergence studies indicate that the relative error grows with both the coupling strength $\Delta$ and the spatial separation $(x_{1}-x_{0})$, necessitating more samples for the intermediate and strong coupling regimes.

\begin{figure*}[h]
    \centering
    \begin{subfigure}[b]{0.45\linewidth}
        \centering
        \includegraphics[width=\linewidth]{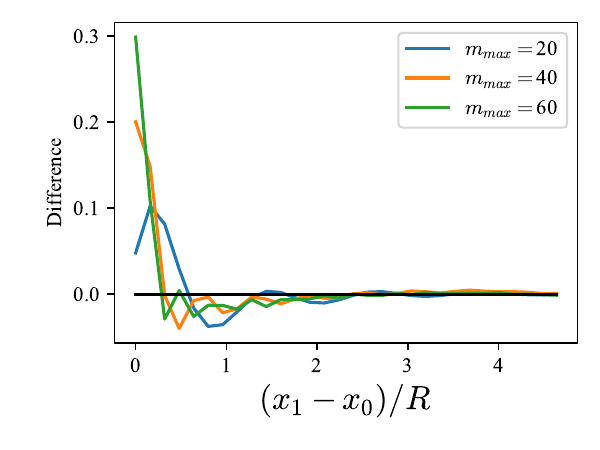}
    \end{subfigure}
    \begin{subfigure}[b]{0.45\linewidth}
        \centering
        \includegraphics[width=\linewidth]{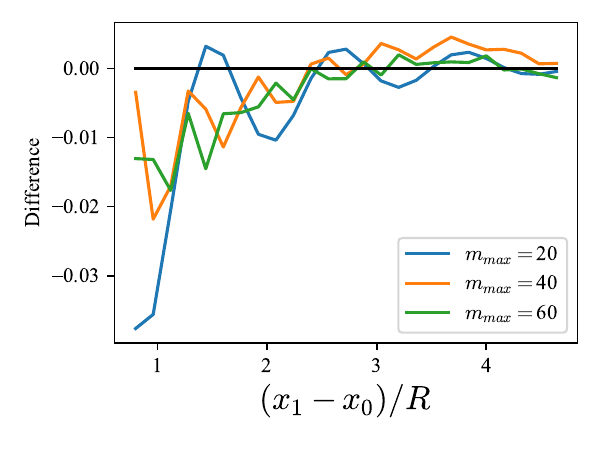}
    \end{subfigure}
    \caption{Analysis of numerical artifacts arising from Fourier mode truncation. The plots show the difference between the connected two-point correlation function and an exponential fit performed on the data with the highest mode cut-off. The left panel displays the full observed spatial range, while the right panel provides a zoomed-in view of the tail. Because the underlying correlation function is not inherently exponential, a larger $m_\text{max}$ (e.g., $m_\text{max}=60$) is necessary to provide a smoother exponential decay and a more precise description of short-range physics by reducing oscillatory peaks.}
    \label{fig:mmaxdependence}
\end{figure*}

\begin{figure*}[h]    
    \centering
    \includegraphics[width=\linewidth]{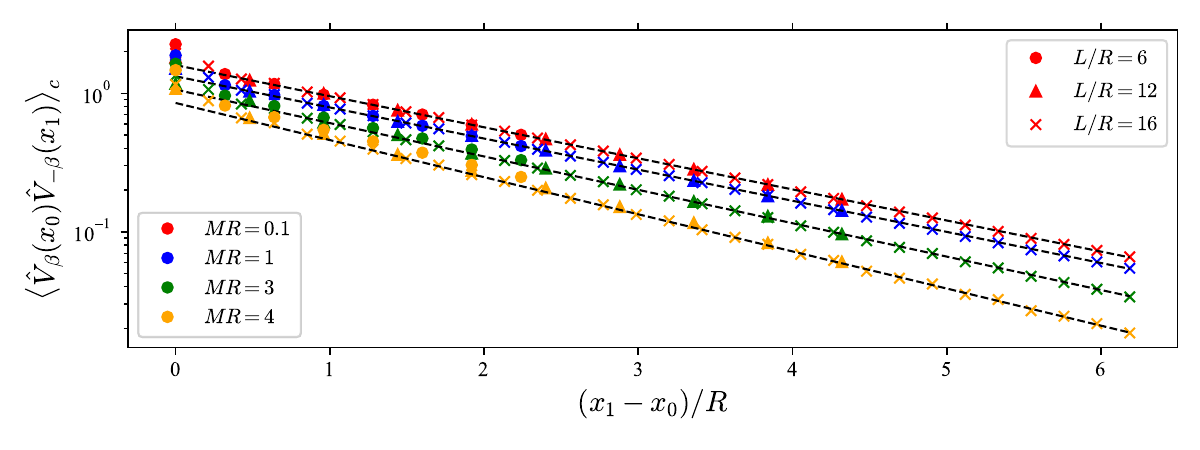}
    \caption{Equal-time correlation function of vertex operators for $\Delta=2/25$ for different inverse temperatures and geometries. The dashed lines represent exponential fits on the tails of the largest $L/R$ results. We observe good agreement for the different $L/R$ ratios, and the correlation lengths seem to be the same for all three. The prefactor of the exponential fits suffer a minor finite size effect which increases with the inverse temperature.\label{fig:FSE}}
\end{figure*}

To implement the MRS, several controlled approximations are introduced:
\begin{itemize}
    \item Fourier mode truncation ($m_\text{max}$): The number of Fourier modes is truncated at $m_\text{max} = 60$ to provide a necessary ultraviolet (UV) cut-off. Using a larger $m_\text{max}$ yields a smoother exponential decay and a more accurate description of short-range correlations. The $m_\text{max}$-dependence is studied in Fig.\ \ref{fig:mmaxdependence}. 
    \item Finite system size ($L$): We utilize discrete Fourier expansions by restricting the spatial integration to the interval $[-L/2, L/2]$ and enforcing periodicity on the Green's function.
    \item Grid discretization: The basis functions $\psi_\ff$ are evaluated on a discrete grid for numerical integration of the $g(\{t_\ff\})$ terms.
\end{itemize}
Stability of results is ensured by benchmarking across different geometries ($L/R = 6, 12, 16$) (see Fig.\ \ref{fig:FSE}). While the previous study \cite{Toth2025} found a noticeable finite-size effect in the expectation value and the free energy density, this effect appears to be very small for the correlation length. At low temperatures, there is some numerical uncertainty, mainly due to larger numerical errors that degrade the quality of the exponential fits. However, in the high-temperature regime, where numerical errors are better controlled, the correlation length is essentially independent of the system size.

\end{document}